# Modified Minimum Connected Dominating Set formation for Wireless Adhoc Networks

Mano Yadav, Vinay Rishiwal, G. Arora and S. Makka

**Abstract-** Nodes of minimum connected dominating set (MCDS) form a virtual backbone in a wireless adhoc network. In this paper, a modified approach is presented to determine MCDS of an underlying graph of a Wireless Adhoc network. Simulation results for a variety of graphs indicate that the approach is efficient in determining the MCDS as compared to other existing techniques.

——————————  ◆  ——————————

## 1 INTRODUCTION

MANET is an infrastructure less multihop network. Routing in MANETs is a one of the challenging task to accomplish. To generate routes proactively or on-demand is extremely costly for energy constrained nodes in a limited bandwidth shared wireless channel. Most of the protocols of MANETs [1] make use of blind broadcast at the time of route discovery whether it is done proactively or in a reactive manner. Communication by blind broadcast induces an intolerable overhead is not a feasible solution. A backbone similar to wired infrastructure network is required for cost effective communication and maintenance of the route. It is therefore; proposed to restrict the routing process in wireless ad hoc networks thereby, to the formation of a virtual backbone using MCDS approach. A virtual backbone reduces the communication overhead, increase the bandwidth efficiency, reduce channel bandwidth consumption, decrease the energy consumption and increases network operational life. A minimum connected dominating set (MCDS) [2, 6, 7] can be optimal virtual backbone in such networks. However, determination of MCDS in a graph is an NP-hard problem [8].

The rest of the paper is organized as follows. Section 2, presents the back ground details and previous study for finding a MCDS. Section 3 contains the details of algorithm1(ModifiedMCDS)and algorithm2(MCDS2). In Section 4, the Modified algorithm is evaluated and compared to other existing approach. In the last section, the work is concluded.

## 2. BACKGROUND AND PREVIOUS STUDY

A connected dominating set (CDS) can be used to reduce redundancy due to blind broadcasts. In a simple graph G (N, E), N is the set of nodes and E is the set of edges. Assume a node set T subset of N such that for all u in N -T, there exists v € T, such that edge (u, v) € E. This is the cover property for the CDS. Set T is called a dominating set. Set T is also called a connected dominating set (CDS)

- *Vinay Risiwal is with the MJP Rohilkhand University, Bareilly, UP, India.*
- *Mano Yadav, G. Arora, and S. Makka are with the ITS Engineering College, Greater Noida, UP, India.*

when T forms a connected graph. This is the connectivity property for the CDS.

Figure 1 gives an example of a CDS. Nodes inside the circles are connected and cover all nodes in the network. They form a CDS for this graph. It is obvious that broadcasting by using nodes in a CDS as relay nodes can reduce redundant rebroadcasts compared with blind broadcast. Broadcast messages can be propagated to all nodes in the CDS because of the connectivity property of the CDS. Non-CDS nodes are able to receive messages because of the cover property of the CDS.

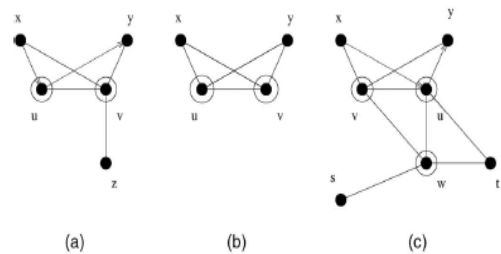

**Figure 1:** An example of a CDS in a simple network.

Broadcasting based on a small CDS can be a promising approach to reduce routing overhead.

### 2.1 Previous Study
A few relevant proposals to CDS formation are as follows:

### 2.1.1 Distributed Approximation Algorithm for Constructing MCDS by Bo Gao & Yuhang Yang [4]

This distributed algorithm to construct approximation MCDS can be briefly described as two phases. In the first phase, a MIS *S* is constructed. The nodes in *S* are referred to as dominators, and the nodes not in *S* are referred to as dominators. In the second phase, each dominator tries to create paths connecting all its two hops away dominators and three hops away dominators by broadcasting a REQUEST DOMI message that includes a life flag (ttl) and a nodes list which will record the IDs of nodes visited by this message. This message is relayed at most two times in the networks, which means it travels at most three hops. When dominators receives a REQUEST DOMI message from a dominator for the first time, it appends its own ID to the nodes list that is included in this message and decreases the ttl value in the message by one and then broadcasts this message. When a dominator receives a REQUEST DOMI message from other dominators, it makes a decision whether to create a path to this dominator according to whether there already exists a shorter path. If there is no existing path between these two dominators, it will create a connecting path by informing the dominators included in the node list of the REQUEST DOMI message to change their states to connectors. Because the decision on creating a path is made by a dominator receiving the REQUEST DOMI message, we can create a unique path between any pair of dominators that depart from each other at most three hops away.

**Messages and Time Complexity**
The distributed algorithm for constructing a CDS has an O(n) time complexity, and O(n) message complexity.

**2.1.2 MCDS algorithm by Bevan Das [6]**

This distributed approximation algorithm, first finds a small dominating set S. After S is determined, the edges (U, dom (u)) form a spanning forest of G. The next stage of Algorithm connects the fragments from the first stage by using a distributed minimum spanning tree (MST) algorithm. C consists of the interior nodes of the resulting spanning tree and the tree edges between these nodes. The first stage consists of rounds of adding nodes to S.
A node u in S is marked: otherwise, U remains unmarked. The number of unmarked neighbors of a node is its effective degree $\delta(u)$. In each round, U gathers
$\delta(v)$ for all v in $N_2(v)$. Node U is added to S if $\delta(u) > \delta(v)$ for all $v \neq u$ in $N_2(u)$; it uses minimum node ID to break ties. Next, Algorithm labels each component formed by the (U, dom(u)) edges with a common label and then connects these components. Before connecting components, we discard edges that connect two nodes in the same component, and then we weight the remaining edges to favor those edges that do not increase the size of C too much. An edge between a node in S and a node not in S adds only one node to C, while an edge between two nodes not in S adds two nodes to C. Therefore, it use the number of endpoints not in S when assigning weights to edges.

The global algorithms proposed by Das, assume that nodes Keep identical copies of the entire topology. (This is always true in a proactive link state routing protocol.) The algorithm, which we call Das CDS algorithm, defines an effective degree for every node. The effective degree of a node is the number of its non-CDS neighbors. The Das CDS I algorithm has two stages. The first stage is responsible for finding a small dominating set S using a loop. At the beginning of iteration in this Loop, nodes are assigned weights – their effective degree. One iteration ends by adding the node with the largest weight to S. The loop in the first stage stops when S covers all nodes. Note that after the first stage, S may contain several disconnected components. In the second stage, the Das CDS algorithm tries to use the minimum number of extra nodes to connect the components of S so that a CDS can be formed. In this stage, links are assigned weights. If a link connects two non-CDS nodes that are in the same component, this link is assigned a weight of infinity. The weight for other links is the number of end points that are not in S. The lightest links are chosen to connect disjointed components. This algorithm ends when a CDS is found.

**Algorithm Complexity**
The time complexity for the Das CDS algorithm is $O((\nu + \gamma)\Delta)$ where $\nu$ is the number of nodes in the network, $\gamma$ is the size of the final CDS, and $\Delta$ is the maximum node degree. A metric, the performance ratio, is defined to characterize the worst-case performance of approximation algorithms, Assume that an algorithm can find a CDS with size C and that the minimum CDS size is $C^*$. The performance ratio is defined as the maximum value of $C/C^*$.

**2.1.3 Global Ripple Algorithm for Minimal CDS [3]**

The global ripple algorithm (GRCDS) assumes all nodes try to maintain identical copies of the entire network topology. This assumption is reasonable when the network mobility is low. When a link changes, the network is in a transient state in which some nodes have different copies of the topology database. We concentrate on low mobility cases for this algorithm at this stage. Here the terms of low mobility is defined as small link connectivity change rates with respect to the frequency to exchange control messages in the CDS algorithms. (This assumption is required for all proactive routing protocols. Otherwise, none of those proactive routing protocol can work properly)
The algorithm runs locally with global topology information. There are two stages. In the first stage, the algorithm broadcasts a zero-payload or "virtual" message starting at the node with the minimum node ID. The initiator of this message is marked as a potential CDS node. When a node broadcasts, all of its one-hop neighbors can receive the broadcast message.
**Algorithm Complexity:** $O((\gamma + \Delta)\gamma)$

**2.1.4 MCDS Algorithm by Wu & Li [7]**

The Wu-Li algorithm is implemented in a distributed manner. This algorithm uses two phases and assumes that all nodes know all the other nodes that are within their two-hop range. In the first stage, a node is selected as a potential member of the CDS if and only if it has two non-adjacent neighbors. Nodes broadcast if they elect themselves as members of the potential CDS in the first phase. Two extensions are used in the second phase to reduce the size of the CDS. A node stays in the CDS unless a neighbor CDS node with a larger ID covers its entire neighbor set. As an extension, if the neighbor set of a node is covered by two adjacent CDS neighbors with larger IDs, this node may change itself to a non-CDS node.

**Algorithm Complexity**
The performance can be measured by computation and communication complexity. In this approach, the time complexity of the marking process at each vertex is $O(\Delta^2)$. The total amount of message exchanges is $O(\Delta v)$, where $v = |V|$ is the total number of vertices in G. More precisely, the marking process without using two rules needs one round, with applying two rules needs two rounds.

## 3 PROPOSED ALGORITHMS

### 3.1 Algorithm1 (Modified MCDS)
This algorithm is a modification of the Bevan algorithm [6]. The algorithm works in three phases. In the first phase a small dominating set S is determine. A dominating set S of G(V,E) is defined as the set of all nodes in which all nodes are either in S or adjacent to some node v in S.
In the second step algorithm connect all the nodes in S to form connected dominating set (CDS), and in the final third step cover checking and connectivity checking is applied to reduce CDS into minimum connected dominating set.

**Step1. Construction of dominating set (S)**
1. Initially S is empty; a node in S is marked otherwise unmarked.
2. The no of unmarked neighbors of a node is its effective degree $\delta(u)$.
3. In each round u gather $\delta(v)$ for all v in $N_2(u)$ i.e. for all two hop neighbor.
4. Node u will be in S if and only if
   $\delta(u) > \delta(v)$ for all v in $N_2(u)$

**Step2. Construction of connected dominating set**
In this phase, each dominator generates a REQUEST DOMI message to send all other dominators within its three hops. This message is broadcasted at most three times before it arrives at a dominator. Every dominatee appends its ID into the nodes list included in the REQUEST DOMI message when it receives this message and then broadcasts this message. In this way, when a REQUEST DOMI arrives at a dominator, it has already recorded the IDs of all nodes in its nodes list which forms the path from the dominator originating this message to the dominator receiving this message. When a dominator receives a REQUEST DOMI message for the first time from another dominator, it generates a REPLY DOMI message including the path that this message should visit in this message and sends it. This path is the reverse order of the one in the REQUEST DOMI message that it has received before. When the dominatee who is included in the path of the REPLY DOMI message receives this REPLY DOMI message, it changes its state to connector and sends this message to the next-hop node according the path in this message

**Step3 Construction of minimal connected dominating set**
In the final stage, a potential CDS node decides whether or not to stay in the CDS based on the list of potential CDS neighbors, together with neighbor list information. Potential CDS nodes are examined in ascending order of the node ID. A potential CDS node performs a coverage check and a connectivity check in this stage.
**Cover check:** To find out if it's potential CDS neighbors cover its non CDS neighbor.
**Connectivity Check:** To find out Whether or not these CDS neighbors are connected is determined by checking connectivity.

The above proposed algorithm is similar to Bevan algorithm in the way dominating set is constructed, but differ in the way MCDS is constructed. Instead of constructing the minimum spanning tree (having the complexity m log n where m is the no of edges and n is the no of nodes) and taking the interior nodes of MST as CDS, we have adopted the concept of cover checking which has the time complexity of $O(\Delta^2)$ where $\Delta$ is the maximum node degree and the connectivity check time is proportional to $O(\Delta^2)$.

### 3.2 Algorithm2 (MCDS2)

This algorithm directly gives the MCDS nodes for the entire network so it is better in performance than other algorithms.
This algorithm works in two steps.

**Step 1:** Every node has a unique id and the algorithm starts with minimum id node.

**Step 2:** A node i is called the MCDS node when all the neighbors of node i is not covered by neighbors of one or more neighbor of node i.

## 4 SIMULATION FRAMEWORK AND RESULT

In this section, simulation is done which compares the average size of the dominating set generated using modified MCDS approach (MMCDS) and algorithm2 (MCDS2) with the MCDS1 approach in [7]. Random graphs are generated in square units of a 2-D simulation area and a random number of nodes are placed on this area. Links are bidirectional and two nodes are connected if they are in the wireless range of each other. Fig. 2 shows the number of MCDS nodes in generated graphs.

We compare modified approach by varying the number of node, n, on a graph. For each value of n, we compute the size of MCDS. Graphs have been generated for dense topology.

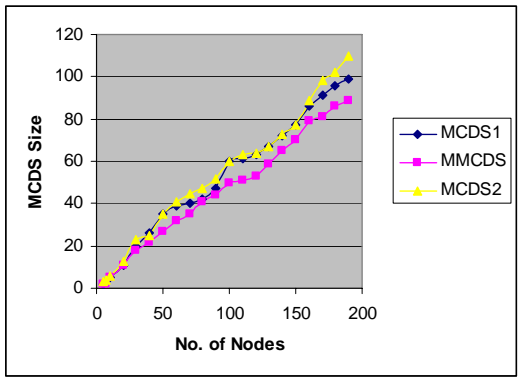

Figure 2: MCDS size versus number of nodes

Simulation shows the performance of MMCDS is better compared to MCDS1 approach [7] and MCDS2 as the network is growing to large size network. MCDS1 and MCDS2 are almost similar in performance for small networks while MCDS1 works better in compare to MCDS2 for large networks. Complexity analysis shows that MCDS2 has less time complexity in worst cases in compare to MCDS1.Figure 2 shows that there is a significant reduction in the size of the MCDS between Modified and existing approach.

## 5 CONCLUSION

In this paper, we have presented a modified algorithm for calculating minimum connected dominating set in the wireless ad hoc network. Complexities of the existing and proposed MCDS approaches are given in this paper. A comparison of the proposed approaches is made with the existing MCDS approach. Simulation result shows that the modified approach outperforms with the existing approach for large networks.

## REFERENCES

[1] Elizabeth Royer and C.K.Toh, "A Review of Current Routing Protocols for Ad-Hoc Mobile Wireless Networks", IEEE Personal CommunicationMagazine, April 1999, pp.46-55.

[2] P.J. Wan, K. M. Alzoubi, O. Frieder, "Distributed Construction of Connected Dominating Set in Wireless Ad Hoc Networks," in IEEE INFOCOM 2002, Pages 1597-1604.

[3] Tao Lin et al's, "Minimum Connected Dominating Set Algorithms and Application for a MANET Routing Protocol," in procedings of the international performance computing and communication conference (IPCCC'03), pp 157-164, phoenix,AZ,April 2003.

[4] Gao et al.'s, " An Effective Distributed Approximation Algorithm For Constructing Dominating Set In Wireless Ad Hoc Networks," proc. of the fourth international conference on computer and information technology(CIT'04),2004.

[5] C. R. Lin and M. Gerla, "Asynchronous multimedia multi-hop wireless networks", manuscript, Computer Science Department, University of California, Los Angeles, 1996, chunghung@cs.ucla.edu, gerla@.cs.ucla.edu.

[6] V. Bharghavan and B. Das, "Routing in Ad Hoc Networks Using Minimum Connected Dominating Set", in Proceedings of International Conference on Communications'97, Montreal, Canada. June 1997.

[7] Wu and H. L. Li, "On Calculating Connected Dominating Set for Efficient Routing in Ad Hoc Wireless Networks", in Proceedings of the 3rd ACM International Workshop on Discrete Algorithms and Methods for Mobile Computing and Communications, 1999, Pages 7-14.

[8] D. B. West, Introduction to Graph Theory, 2nd ed., Prentice Hall, Upper Saddle River, NJ, 2001, pp. 116-118.